\documentclass[global,twocolumn,
]{svjour}

\usepackage{latexsym}
\usepackage{graphics}
\usepackage{graphicx}
\usepackage{amsmath}
\usepackage{amssymb}
\usepackage{bm}

\usepackage[normalem]{ulem} 
\usepackage {ulem}

\journalname{}
\begin{document}
\title{A laser system for the excitation of rubidium Rydberg states using second harmonic generation in a PPLN waveguide crystal}
\titlerunning{A laser system for the excitation of rubidium Rydberg states using SHG in a PPLN waveguide crystal}
\author{Andreas Koglbauer\inst{1} \and Peter W{\"u}rtz\inst{2} \and Tatjana Gericke\inst{2} \and Herwig Ott\inst{2}
}                     
\mail{koglbauer@uni-mainz.de}

\institute{Institut f{\"u}r Physik, Johannes Gutenberg-Universit{\"a}t, Staudingerweg 7, 55128 Mainz, Germany, fax: +496131/39-23428 \and Research Center OPTIMAS, Technische Universit{\"a}t Kaiserslautern, Erwin-Schr{\"o}dinger-Str. 46, 67663 Kaiserslautern, Germany, fax: +49631/205-3906}
\date{Received: date / Revised version: date}
%
\maketitle
\begin{abstract}
We report on a laser system at a wavelength of 495\,nm which is suitable for the excitations of low lying Rydberg states of rubidium atoms. The system is based on frequency doubling of a seeded diode laser in a periodically poled waveguide crystal. We achieve an output power of up to 35\,mW and prove the single frequency performance by direct two photon laser spectroscopy of the rubidium $14D_{5/2}$ and $14D_{3/2}$ states. The measured fine structure splitting is consistent with quantum defect theory calculations.
\end{abstract}
\section{Introduction}

The last years have seen a rapidly increasing interest in Rydberg atoms, driven by the ongoing progress in cold atom physics. The experimental observation of the Rydberg blockade in thermal and condensed samples \cite{Heidemann2007,Reetz-Lamour2008,Heidemann2008} as well as the exploitation of the blockade for quantum logical operations \cite{Urban2009,Gaetan2009} mark the most important breakthroughs. These new developments complement previous studies on cold plasmas and spectroscopy of Rydberg states and are accompanied by exciting proposals for quantum information and quantum simulation \cite{Jaksch2000,Lukin2002,Saffman2010}. 

Most experiments are realized with rubidium atoms where a convenient excitation scheme exists using a two photon transition via the intermediate $5P_{3/2}$ state. The first excitation  step at 780\,nm can be made with a standard diode laser while the second step $\sim480$\,nm can only be realized using second harmonic generation (SHG). An amplified diode laser system (MOPA) with subsequent frequency doubling in a resonator is a frequently used configuration which can be homemade or purchased commercially. In this configuration, the power of the blue light can exceed 200 mW \cite{Toptica}. The bandwidth of the frequency doubled light is determined by that of the fundamental frequency and typically amounts to a few MHz. The approach is cost intensive, and the handling of SHG in a resonator can be involved. Less sophisticated and more cost efficient ways of frequency doubling can therefore be an interesting alternative, especially if only low power at the fundamental frequency is available.

Periodically poled crystals can double the frequency in a single pass. Guiding the fundamental light field in a waveguide provides high intensity, while the periodic poling provides the condition of quasi phase matching. Highly efficient single pass frequency doubling using periodically poled lithium niobate (PPLN) waveguide crystal at 488\,nm has been previously reported \cite{jechow07}. Coupling light from a DFB laser (total output power 400\,mW) in the waveguide (300\,mW coupled power), a converted power of 160\,mW could be generated, corresponding to a conversion factor of $52\,\%$. As shown recently \cite{Kim2010}, this approach is also suitable for narrow linewidth SHG in the kHz regime.

Here, we report on frequency doubling of a seeded diode laser with an output power of 300\,mW and a wavelength of 990\,nm. We achieve up to 35\,mW power at 495\,nm.
 The single-frequency operation of the laser system is verified via EIT spectroscopy of the $5S_{1/2}\rightarrow5P_{3/2}\rightarrow14D_{3/2}/14D_{5/2}$ two photon transition of $^{87}$Rb, thus directly proving its applicability for the excitation of Rydberg states. We find an upper bound for the line\-width of 3\,MHz.

The choice of the $14D$ state of rubidium is motivated by our experimental approach to study ultracold quantum gases with the help of scanning electron microscopy \cite{Gericke2008,Wuertz2009}. In the experiment, the atoms are prepared in a CO$_2$ optical dipole trap. When excited to the 14$D$ state, the atoms have a high probability to be photoionized by the CO$_2$ laser \cite{potvliege06}. This helps to increase the overall detection efficiency of this technique. Details of the ionization scheme can be found in \cite{Markert2010}.

\section{Laser setup}

The laser system for the fundamental light at 990.6\,nm wavelength is a master-slave setup, consisting of two identical high-power laser diodes, with a nominal output power of 400\,mW and a gain bandwidth from 985\,nm to 995\,nm. Both diodes are temperature controlled, and collimated by aspheric lenses. The wavelength selection of the master laser is accomplished by means of optical feedback from a diffraction grating in an external cavity setup in Littman configuration \cite{Liu81}. This provides the advantage of beam pointing stability during wavelength changes, which is crucial for the following injection-locking \cite{hadley86}. As a consequence, in our setup only 16\,\% of the free running power remain after transmission through a 60 dB Faraday isolator. Single-mode operation over the whole gain bandwidth is possible.

\begin{figure}[htbp]
  \includegraphics[width=8.3cm]{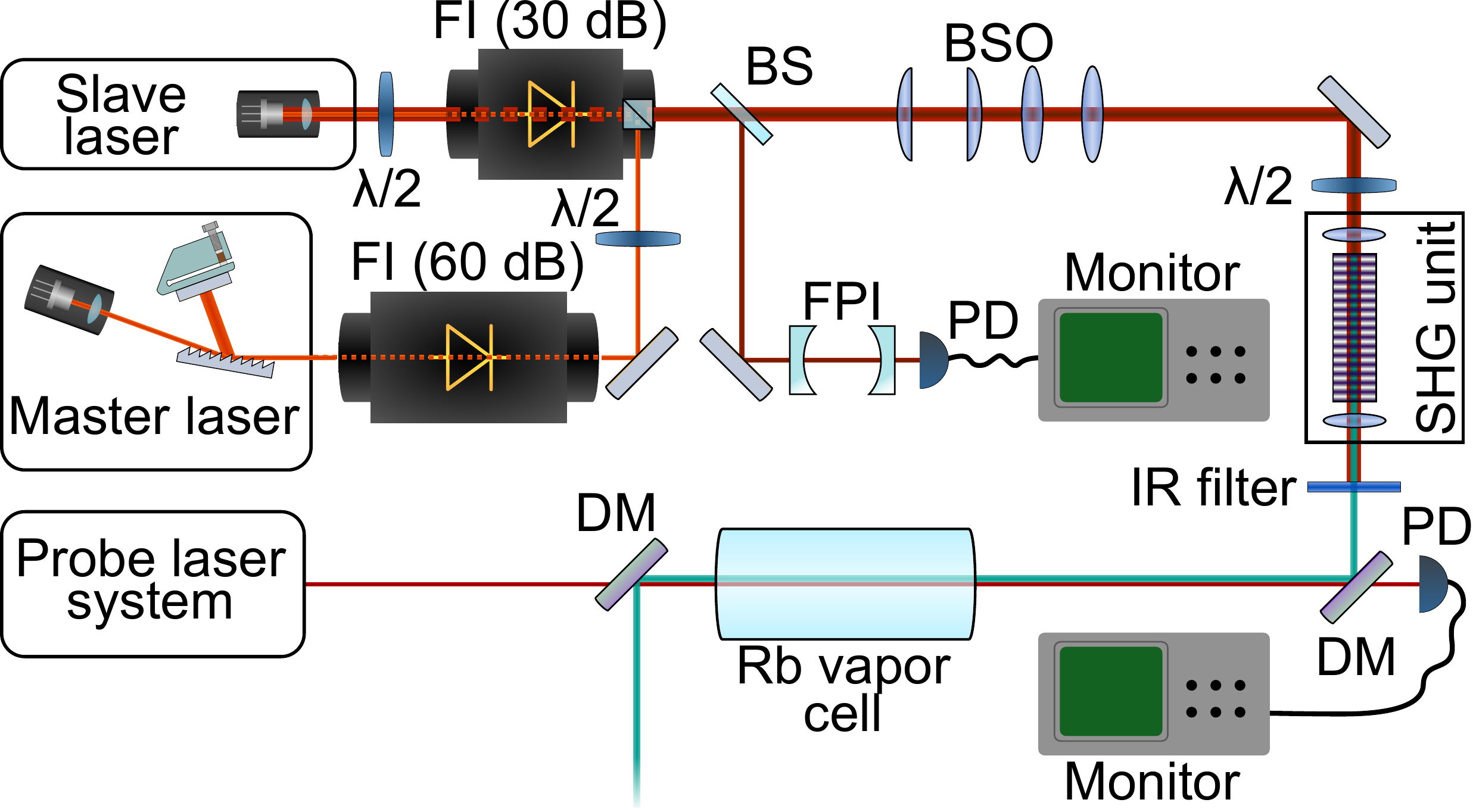}
\caption{Sketch of the laser setup. The master diode in Littman configuration seeds the slave laser, whose frequency locking is monitored via a spectral analyzer. After beam shaping optics (BSO) the light is coupled into the PPLN waveguide for SHG. Residual fundamental light is blocked by an IR filter. The blue coupling beam at 495 nm passes through a Rb vapor cell and is aligned with the probe beam at 780 nm. The latter is provided by an offset-lock stabilized diode laser system. The transmission of the probe beam is observed by means of a photo diode behind a dichroic mirror. Where necessary, polarization adjusting is performed via half-wave plates ($\lambda/2$). (FI: Faraday isolator, BS: Beam splitter, FPI: Fabry P{\'e}rot interferometer, PD: Photo diode, DM: Dichroic mirror)}
\label{fig:laser_setup}
\end{figure}

The complete setup is depicted in Fig. \ref{fig:laser_setup}. In order to couple the master beam into the front facet of the slave diode, it is overlapped with the beam of the slave laser, using the side entrance of the output PBC in a 30 dB Faraday isolator. Since the two diodes are identical in construction, no further mode matching is required. Typically about 23\,mW of injection light are available at this point. 
This is sufficient to seed the slave laser up to the maximal extracted output power of 370\,mW for several hours, if the master laser doesn't experience any mode hopping exceeding the locking range. Frequency scans of $>1$\,GHz are supported by the injection lock. A reduction of the injection power to $\mathord<1$\,mW, while maintaining the maximum slave power is possible, though the frequency locking range decreases. The inspection of single-mode operation is accomplished via spectral analysis of a small fraction of the beam using a Fabry P{\'e}rot interferometer (FPI) and a photo diode. After beam shaping optics to compensate for the strongly elliptical profile of the diode and adjustment of the spot size, the infrared light is coupled into the frequency doubling setup. The polarization of the laser is matched to the crystal axis via a half-wave plate prior to the SHG unit, where infrared powers of 300\,mW are available for conversion. The frequency conversion assembly consists of two aspheric lenses for the coupling of the incident fundamental ($f\mathord=3.3$\,mm, $NA\mathord=0.47$) and collimation of the emitted harmonic light ($f\mathord=5.0$\,mm, $NA\mathord=0.55$), and a temperature controlled mounting for the waveguide crystal. The latter provides a temperature stability of $0.1$ K. Lenses and crystal can be positioned relative to each other by two three-axis translation stages. The blue beam is separated from the non-converted transmitted fundamental light through an edge filter.

\section{Frequency conversion in a PPLN waveguide crystal}

We use as nonlinear medium a periodically poled magnesium oxide doped lithium niobate waveguide crystal (HCPhotonics). It has outer dimensions of $10\times3.6\times0.5$\,mm$^3$ and is designed for the SHG of infrared light in the range of $\lambda=990-997$\,nm at quasi phase matching temperatures between 100$^\circ$C and 170$^\circ$C with a normalized conversion efficiency $\eta=150-200$\,\%W$^{-1}$cm$^{-2}$. The input and output facet of the crystal are anti  reflection coated for the fundamental and harmonic wavelength. On the top face is a total of 36 waveguide channels with a rectangular transverse profile of a few $\mu$m$^2$. Whereas only one channel within the specifications is guaranteed by the manufacturer, there is always a small amount of conversion in every channel. Hence, the working channel can be easily identified by horizontal translation of the crystal in the focal plane of the focussing lens and counting of the waveguides.

In order to measure the coupling efficiency into the waveguide, the crystal temperature is detuned far from the optimal phase matching temperature, so that attenuation of the infrared light by SHG can be neglected.  A coupling efficiency of 68\,\% could be achieved, corresponding to a maximum of 200\,mW within the crystal. With light originating from a single mode fiber even higher coupling efficiencies ($>80$\,\%) are feasible \cite{HCP}.

For a lossless waveguide the theoretically obtainable power $P_{2\omega}$ of the harmonic light field for perfect phase matching $\Delta k_Q = 0$ after an interaction length $z$ yields \cite{parameswaran02}:
	\begin{equation}
	P_{2 \omega}=P_{\omega} \tanh^2\left[\sqrt{\eta P_{\omega} z^2}\right]\ .
	\label{eq:ISHG-strong}
	\end{equation}   
Here $P_{\omega}$ is the power of the incident fundamental light, and $\eta$ the normalized conversion efficiency. Due to the temperature and wavelength dependency of the refractive index as well as the thermal expansion of the crystal, the condition $\Delta k_Q = 0$ can be met by an appropriate choice of one of these two parameters, if the other one is fixed. 

\begin{figure}[htbp]
  \includegraphics[width=8.3cm]{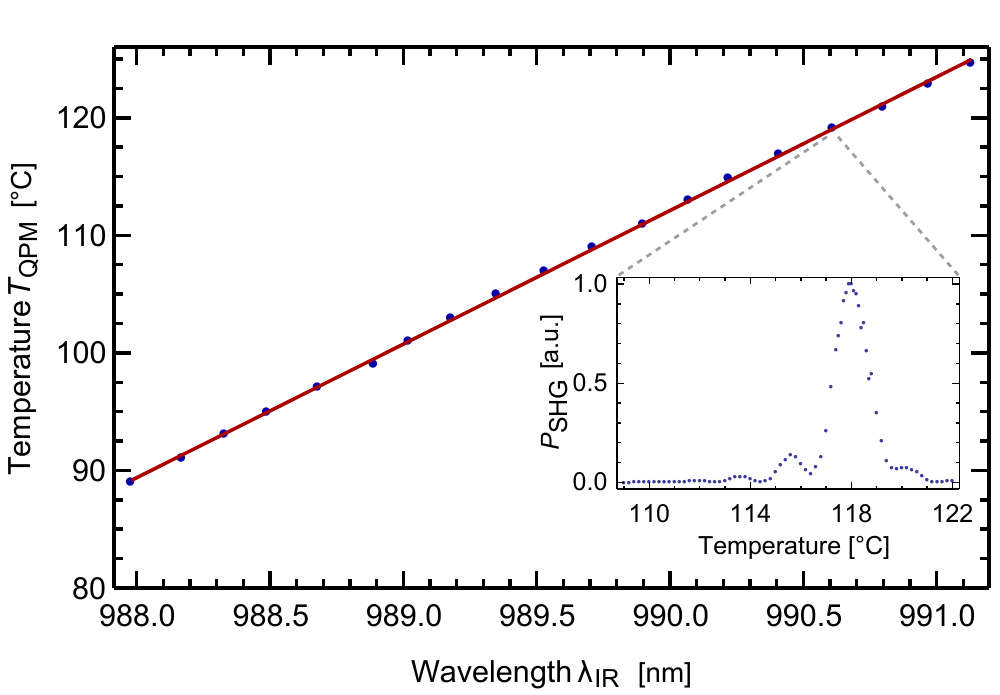}
\caption{Tuning range of the blue laser. Temperature of optimal phase matching as a function of the fundamental wavelength over a range of 3\,nm (blue data points). A linear regression with a slope of $11.4$\,K/nm reproduces the data very well (red line). The phase matching curve for a variation of the temperature, with a fundamental light field of $P_{\text{IR}}=65.6$\,mW at 990.6\,nm is shown in the inset. The graph is normalized to its peak value.}
\label{fig:temperature_dependence}
\end{figure}

Simultaneous variation of crystal temperature and fundamental wavelength allows for a tuning of the blue laser over several nanometers. In Fig. \ref{fig:temperature_dependence} the temperature for optimal quasi phase matching is shown as a function of the wavelength of the infrared light over a range of 3\,nm. A linear behavior with a slope of $11.4$ K/nm is apparent, demonstrating the tuning abilities of the system.  The measured relative intensity of the blue light at a fixed wavelength as a function of the temperature is given in the inset of Fig. \ref{fig:temperature_dependence} for 990.6\,nm.

\begin{figure}[htbp]
  \includegraphics[width=8.3cm]{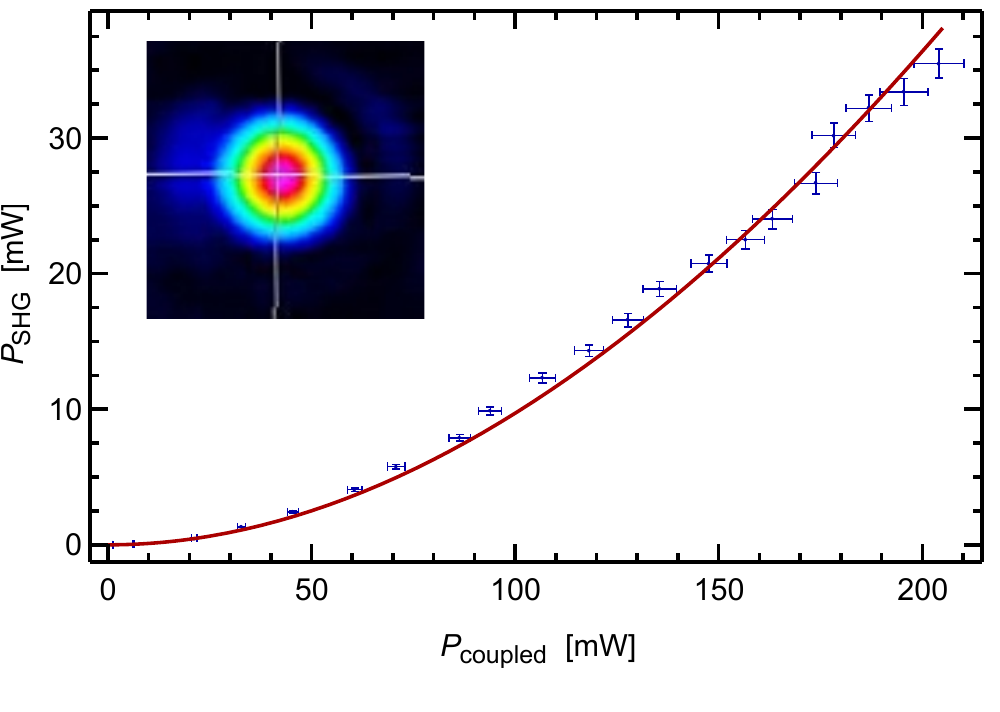}
\caption{Blue laser power as a function of the fundamental power within the waveguide channel (blue data points). Errorbars are due to measurement inaccuracies. The theoretical curve (following equation~(\ref{eq:ISHG-strong})) yields a normalized conversion efficiency of $\eta=103$\,\%W$^{-1}$cm$^{-2}$ (red curve). A maximal power of $P_{\text{SHG}}=35.5$\,mW was achieved. The inset shows a beam profile of the frequency doubled light.}
\label{fig:conversion_efficiency}
\end{figure}

The maximal achieved optical output power at $495.3$ nm wavelength with this system is $35.5$ mW, converted from 204 mW infrared light within the crystal. This corresponds to a conversion efficiency of  17.4\,\% if coupling losses are not considered. Including latter, the conversion efficiency amounts to 11.8\,\%. Fig. \ref{fig:conversion_efficiency} presents the corresponding data. A theory curve following (\ref{eq:ISHG-strong}) with a normalised conversion efficiency of $\eta=103$\,\%W$^{-1}$cm$^{-2}$ reproduces the data. 

\begin{figure}[htbp]
  \includegraphics[width=8.3cm]{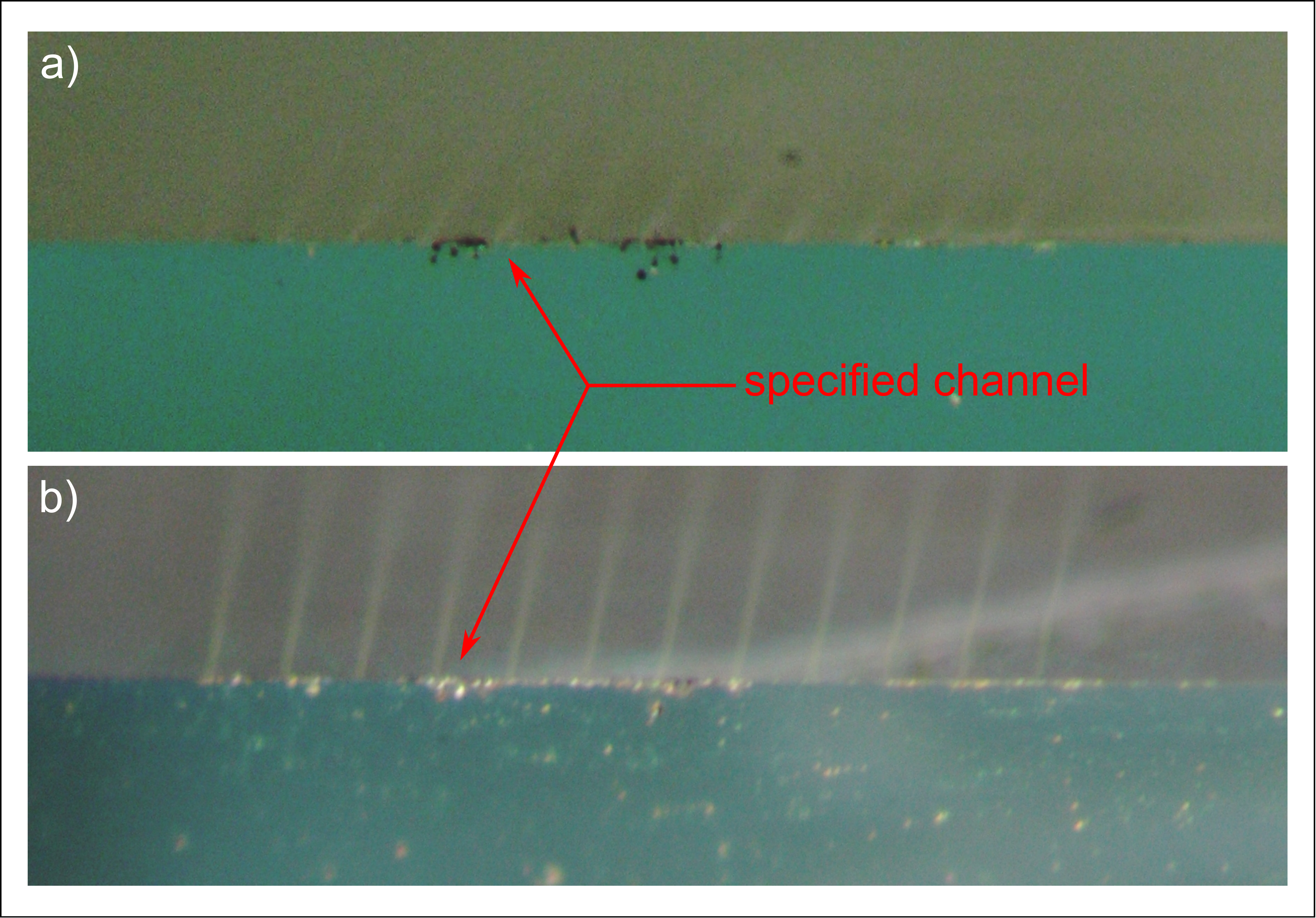}
\caption{Microscopic images of the crystal front facet. a) Accumulation of dust particles near the top edge. b) After cleaning surface damages due to high intensity exposure are visible.}

\label{fig:crystal}
\end{figure}

The transverse mode profile of the frequency doubled light (inset of Fig. \ref{fig:conversion_efficiency}) shows a gaussian distribution in both axes. Injection locking as well as SHG show a good long-time performance suitable for atomic physics experiments \cite{Markert2010}, with only a slight decrease in the blue power due to misalignment. This drift is caused by slow thermal expansion of the crystal oven and can be compensated for. 
However, the achieved normalized conversion efficiency is distinctly lower than in the specifications, and looking at the long-term performance of the crystal, we found a degradation of the waveguide channel. After several months, the conversion efficiency has dropped by more than 50\,\%.  An inspection of the crystal front facet with an optical microscope reveals an accumulation of dust particles in the vicinity of the waveguide channels (dark spots in Fig. \ref{fig:crystal} a)). Cleaning with inert dusting gas can remove latter, but irreversible surface damage remains (see Fig. \ref{fig:crystal} b)). A pigtailed waveguide crystal or a sealed housing could circumvent this problem.

\section{Spectroscopy of Rydberg states}

The wavelength of $495.3$\,nm corresponds to the $5P_{3/2}\rightarrow14D$ transition in $^{87}$Rb which can be calculated by quantum defect theory \cite{Lorenzen83}. The experimental confirmation of the wavelength is performed by a two photon spectroscopy, based on electromagnetically induced transparency, involving the transition in question. Moreover the width of the EIT signal allows an estimate  for the laser linewidth.
The concept of EIT is widely studied \cite{fleischhauer05} and can be understood as an interference of the excitation pathways of atomic states, whose coherence is induced by the laser fields. It manifests as an absence of absorption of an atomic probe resonance, due to the presence of a coupling light field and has already been demonstrated in the ladder excitation scheme of rubidium for lower and higher principle quantum numbers \cite{clarke01,mohapatra07}.

The probe beam laser system consists of two self constructed grating stabilized diode lasers in Littrow configuration~\cite{ricci95} with a wavelength of 780\,nm. The frequency of a first reference laser is fixed to the atomic $(1,3)$-crossover peak of the $5S_{1/2},F\mathord=2 \rightarrow 5P_{3/2}$ resonance in $^{87}$Rb by a doppler-free FM-spectroscopy~\cite{bjorklund80} in a vapor cell. Beating with a second reference laser to establish an offset lock enables us to shift the frequency of the reference laser 2 up to 1.5\,GHz with respect to the atomic transition. Hence, the system provides a tunable probe laser beam for the EIT-spectroscopy and is aligned with the blue coupling beam in counter propagating direction through a rubidium vapor cell at room temperature (see also Fig. \ref{fig:laser_setup}). The absorption signal of the probe beam is detected via a photo diode behind a dichroic mirror, which is transmissive for 780\,nm.

\begin{figure}[htbp]
  \includegraphics[width=8.3cm]{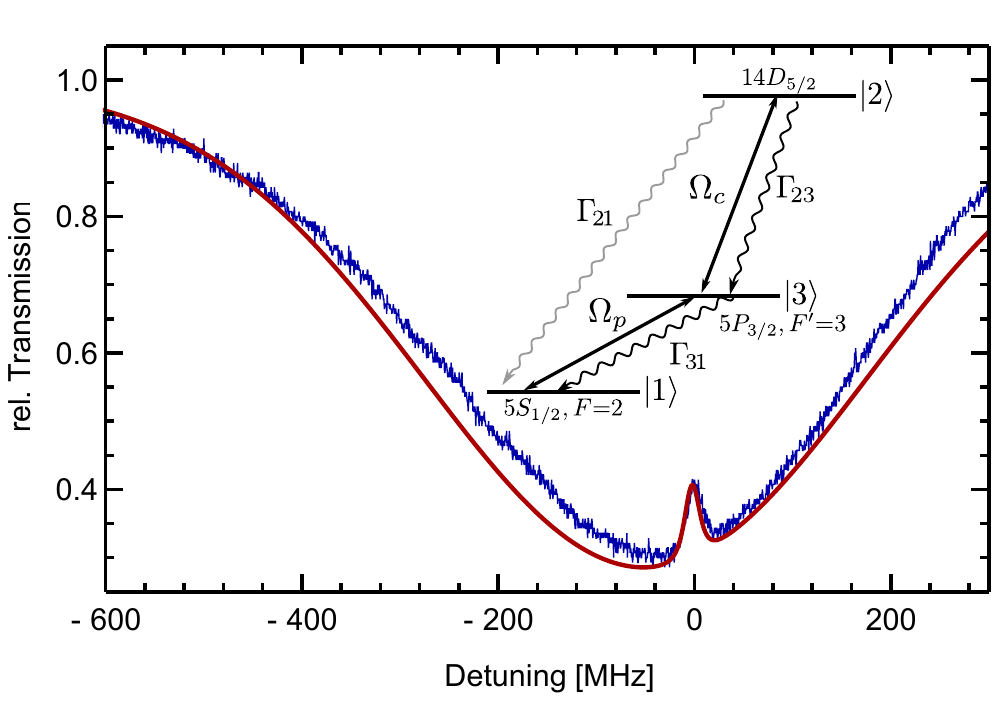}
\caption{Measured normalized transmission spectrum of the probe beam in the presence of the coupling beam tuned close to the  $5P_{3/2},F'\mathord=3\rightarrow14D_{5/2}$ transition, and theoretical lineshape for Rabi frequencies $\Omega_p=12.3$\,MHz, and $\Omega_c=2.3$\,MHz. The probe beam was scanned over the doppler-broaded $D_2$-line in $^{87}$Rb. Theoretical and experimental EIT-signals have a FWHM of $\sigma_{\textrm{theo}}\mathord=14.2$\,MHz and $\sigma_{\textrm{exp}}\mathord=14.6$\,MHz and amount to a reduction of the absorption of 15.4\,\% and 14.4\,\%, respectively. The inset shows the relevant levels and transition for the OBE.}

\label{fig:EIT-signal}
\end{figure}

For a theoretical calculation of the EIT line shape, we solve the optical Bloch equations (OBE) for a three level system, taking into account the thermal motion of the atoms as presented in \cite{beyer09}. For simplification we only considered the $F'\mathord=3$ hyperfine state of the $5P_{3/2}$ level. The equations read as follows 
\begin{subequations}\label{eqn:binomi}
  \begin{align}
  \dot{\rho}_{11}=  i&\Omega_p(\tilde{\rho}_{31}-\tilde{\rho}_{13})+\Gamma_{31}\rho_{33}+\Gamma_{21}\rho_{22}, \label{subeqn-1:binomi} \\
  \dot{\rho}_{33}=  i&\Omega_p(\tilde{\rho}_{13}-\tilde{\rho}_{31})+i\Omega_c(\tilde{\rho}_{23}-\tilde{\rho}_{32})\nonumber\\
  								  &+\Gamma_{23}\rho_{22}-\Gamma_{31}\rho_{33}, \label{subeqn-2:binomi} \\
  \dot{\rho}_{22}=  i&\Omega_c(\tilde{\rho}_{32}-\tilde{\rho}_{23})-(\Gamma_{21}+\Gamma_{23})\rho_{22}, \label{subeqn-3:binomi} \\
  \dot{\tilde{\rho}}_{13}= i&\Omega_p(\rho_{33}-\rho_{11})-i\Omega_c\tilde{\rho}_{12}-(i\delta_{13}+\gamma_{31})\tilde{\rho}_{13}, \label{subeqn-4:binomi} \\
  \dot{\tilde{\rho}}_{23}= i&\Omega_c(\rho_{33}-\rho_{22})-i\Omega_p\tilde{\rho}_{21}+(i\delta_{23}-\gamma_{23})\tilde{\rho}_{23}, \label{subeqn-5:binomi} \\
  \dot{\tilde{\rho}}_{12}= i&\Omega_p\tilde{\rho}_{32}-i\Omega_c\tilde{\rho}_{13}-(i(\delta_{13}+\delta_{23})+\gamma_{21})\tilde{\rho}_{12}. \label{subeqn-6:binomi}
  \end{align}
\end{subequations}
The states $|1\rangle$, $|2\rangle$ and $|3\rangle$ are the ground, Rydberg and intermediate state, respectively (see the inset of Fig. \ref{fig:EIT-signal}), $\rho_{ii}$ describe the populations in the levels, while $\tilde{\rho_{ij}}$ denote the coherences. The Rabi frequencies of the probe and coupling beam in our case amount to $\Omega_p\mathord=12.3$\,MHz and $\Omega_c\mathord=2.3$\,MHz, the corresponding detunings of the applied lightfields are $\delta_{13}$ and $\delta_{23}$. The Doppler shift is introduced by the substitution $\delta_{ij}\mathord\rightarrow\delta_{ij}+\vec{k_{ij}}\cdot\vec{v}/(2\pi)$ and integration over the thermal velocity distribution. The $5P_{3/2}$ state only decays into the ground state with a decay rate of $\Gamma _{31}=\tau_3^{-1}\mathord=38.12$\,MHz. The lifetime $\tau_2\mathord=$2275 ns of the Rydberg state \cite{theodosiou84} is composed of the decay into the intermediate state and the decay to the ground state via levels, not participating in this excitation scheme. Although the transition $|2\rangle \rightarrow |1\rangle$ is only allowed as a multi photon transition, we can approximate it's contribution by identifing $\Gamma_{21}$ with the multi photon decay rate to the ground state, and thus $\tau_2\mathord \simeq (\Gamma_{23}+\Gamma_{21})^{-1}$. We derive $\Gamma_{23}$ from a calculation of the corresponding  dipole matrix element, involving numerical computation of the radial wave functions of the $5P_{3/2}$ and $14D_{5/2}$ states \cite{blatt67,bhatti81}. The dephasing $\gamma_{ij}$ of the coherences are defined 
\begin{subequations}\label{eqn:dephasing}
  \begin{align}
 		\gamma_{31}=&\frac{1}{2}\Gamma_{31}, \label{subeqn-1:dephasing}\\
		\gamma_{23}=&\frac{1}{2}(\Gamma_{21}+\Gamma_{23}+\Gamma_{31}), \label{subeqn-2:dephasing}\\
		\gamma_{21}=&\frac{1}{2}(\Gamma_{21}+\Gamma_{23})\label{subeqn-3:dephasing},
	\end{align}
\end{subequations}
neglecting collision broadening. The imaginary part of $\tilde{\rho_{13}}$ is than associated with the absorption coefficient on the lower transition.

The theoretical curve for the $D_2$ absorption line for rubidium vapor at room temperature  ($n_{\textrm{Rb}}=1.04\times10^{16}$ m$^{-3}$) is shown in Fig. \ref{fig:EIT-signal} (red solid line) together with the normalized transmission spectrum of the scanned probe beam in the presence of the coupling laser (blue data). Within the absorption dip of the doppler-broaded $5S_{1/2}$, $F\mathord=2\rightarrow5P_{3/2}$ resonance, the EIT signal associated with the $5P_{3/2}$, $F'\mathord=3\rightarrow14D_{5/2}$ Rydberg excitation is clearly visible. The measured FWHM of the lorenzian shaped resonance peak is $\sigma_{\textrm{exp}}\mathord=14.6$\,MHz and was calculated to be $\sigma_{\textrm{theo}}\mathord=14.2$\,MHz, hence an additional broadening of $\mathord<\,3$\,MHz due to the laser linewidth can be estimated. This is compatible to typical ECDL systems. Also the decrease in absorption of theory (15.4\,\%) and  experiment (14.4\,\%) show good agreement. Only the absolute height of the theoretical curve was adjusted to fit the data.

A frequency stabilization of the master laser, alike the one described in \cite{abel09} on this signal is under development.

Although six transitions should appear in our excitation scheme (the hyperfine splitting of the Rydberg states is not resolved), we were only able to detect the two strongest resonances $5P_{3/2},F'\mathord=3\rightarrow14D_{3/2}$ and $5P_{3/2},F'\mathord=3\rightarrow14D_{5/2}$. Transitions via the intermediate $F'=1,2$ hyperfine states were not detectable. With reference to these two transitions, we measured a fine structure splitting of the $14D$ Rydberg state of $5.13(35)$\,GHz which is consistent with quantum defect theory calculations. The uncertainty of this value is dominated by the accuracy of the wavelength measurement.

\section{Conclusion and Outlook}
 
We have described a laser system based on single pass frequency doubling of a seeded diode laser, using a PPLN waveguide crystal. The maximum output power was 35.5 mW at 495\,nm wavelength. The power and the line width are sufficient for a large number of spectroscopic applications and quantum state engineering of low lying Rydberg states in rubidium. The wavelength can easily be tuned over several nanometers changing the temperature of the crystal.

The conversion efficiency can be further increased by using longer waveguide crystals. A sealed housing of the crystal is recommended to ensure a long lifetime. Moreover pigtailed crystal setups may offer an insensitivity  to thermal drifts of the oven. The laser system is cost efficient, comparably easy to align and can be an alternative to high power amplified diode laser systems followed by conventional SHG in a cavity.

\begin{acknowledgement}
We would like to thank K. Wu of HCPhotonics and A. Jechow form the university of Potsdam for valuable discussions. This work was supported by the DFG (OT 222/2-3).
\end{acknowledgement}

\end{document}